\documentclass[prb,twocolumn,showpacs,preprintnumbers,floatfix]{revtex4}
\usepackage{graphicx}
\usepackage{dcolumn}
\usepackage{bm}
\usepackage{latexsym,epsfig}
\usepackage{graphicx}
\usepackage{verbatim}
\usepackage{comment}
\usepackage{amsmath}
\usepackage{amssymb}
\usepackage{stmaryrd}
\usepackage{color}
\usepackage{epstopdf}
\DeclareGraphicsExtensions{.eps}

\newcommand{\beq}{\begin{equation}}
\newcommand{\eeq}{\end{equation}}
\newcommand{\bea}{\begin{eqnarray}}
\newcommand{\eea}{\end{eqnarray}}
\newcommand{\ben}{\begin{eqnarray*}}
\newcommand{\een}{\end{eqnarray*}}
\newcommand{\bfig}{\begin{figure}}
\newcommand{\efig}{\end{figure}}

\begin{document}
\title{Frustration-induced supersolids in the absence of inter-site interactions}
\author{Tapan Mishra}
\affiliation{Institut f\"ur Theoretische Physik, Leibniz Universit\"at Hannover, 30167~Hannover, Germany}
\author {Sebastian Greschner}
\affiliation{Institut f\"ur Theoretische Physik, Leibniz Universit\"at Hannover, 30167~Hannover, Germany}
\author{Luis Santos}
\affiliation{Institut f\"ur Theoretische Physik, Leibniz Universit\"at Hannover, 30167~Hannover, Germany}

\date{\today}

\begin{abstract}
We discuss a novel mechanism for the realization of supersolids in lattices in the absence of inter-site interactions that 
surprisingly works as well at unit filling. This mechanism, that we study for the case of the sawtooth lattice, is based on the existence of 
frustrated and un-frustrated plaquettes. For sufficiently large interactions and frustration the particles gather preferentially 
at un-frustrated plaquettes breaking spontaneously translational invariance, resulting in a supersolid.
We show that for the sawtooth lattice the supersolid exists for a large region of parameters for densities above half-filling.
Our results open a new feasible path for realizing supersolids 
in existing ultra-cold atomic gases in optical lattices without the need of long-range interactions.
 \end{abstract}






\maketitle

\section{Introduction}
Supersolids have attracted a large interest since they were proposed~\cite{Andreev1969,Chester1970}, due to their, 
apparently counter-intuitive, co-existence of both crystalline order and superfluidity.
Superfluidity on top of the crystalline order is explained by the creation and delocalization of  
zero point defects (such as vacancies or interstitials) in a strongly interacting system. 
The search for the elusive supersolid remains a challenge~\cite{Prokofiev2005,Balibar2010,Boninsegni2012}. The claimed evidence of supersolidity 
in Helium~\cite{Kim2004}, was subsequently explained by the shear modulus stiffening of solid $^{4}$He~\cite{Boninsegni2012,Kuklov2011}. 
 
Supersolidity may occur as well in lattices due to inter-site interactions. 
Lattice supersolidity has attracted an active theoretical interest as well~\cite{Baranov2012}. 
Ultra-cold gases in optical lattices provide an interesting system for the realization of lattice supersolids.  
However, the requirement of sufficiently large inter-site interactions reduces the possible scenarios for supersolidity to 
gases with long-range interactions, including dipolar gases~\cite{Baranov2012,Lahaye2009} and Rydberg--dressed atomic gases~\cite{Gallagher2008,Cinti2014}. 
These systems present however  difficulties due to inelastic collisions in polar molecules~\cite{Ospelkaus2010} and short life times in Rydberg gases. 
An alternative to polar gases is provided by experiments with condensates in optical cavities, where 
infinitely long-range interactions between the condensed atoms are induced by two-photon processes. These interactions drive the Dicke phase transition 
that results in self-organized supersolids~\cite{Baumann2010}. Although these experiments are realized in absence of optical lattices, 
the self-organized supersolids break spontaneously a discrete spatial symmetry, hence resembling the case of lattice supersolids.

Long-range interactions are however not necessary for the realization of lattice supersolids. Recent studies on frustrated lattices with flat bands, such as kagome lattices~\cite{Huber2010} and Creutz ladders~\cite{Takayoshi2013,Tovmasyan2013}, 
have discussed the possibility of observing supersolids without inter-site interactions. In the case of the Creutz ladder, 
supersolidity at incommensurate densities results from effective next-to-nearest neighbor hopping in the vicinity of the two flat band region, and the doping of a Valence Bond Crystal.

In this paper we discuss a novel mechanism that leads to robust lattice supersolidity in the absence of inter-site interactions. The mechanism 
is based on the existence of frustrated and un-frustrated plaquettes. In this sense,  although we study the specific case of a sawtooth lattice, 
we expect that supersolids may be also realized in other lattice geometries fulfilling this property.
Supersolidity follows from the preferential occupation of un-frustrated plaquettes, which results in spontaneously broken translational 
symmetry. Sawtooth lattices and other frustrated geometries may be realized using lattice shaking~\cite{Struck2011} and similar laser arrangements as those recently employed for creating 
variable lattice geometries~\cite{Wirth2011,Tarruell2012,Jo2012}. Hence our results open a new feasible path for realizing supersolidity in existing atomic gases 
without the need of long-range interactions.

The structure of the paper is as follows. In Sec.~\ref{sec:Model} we introduce the sawtooth model studied in the paper. Section~\ref{sec:Non-interacting} discusses the non-interacting regime. 
Section~\ref{sec:Ground-state} is devoted to the ground state phase diagram, both at unit-filling and away from unit filling. In Sec.~\ref{sec:Weak-coupling} we introduce a simple model 
that allows an intuitive understanding of the super solid mechanism. Section~\ref{sec:Roton} discusses the roton instability responsible for the superfluid-to-supersolid transition. In Sec.~\ref{sec:Experimental} 
we comment on the experimental realization of the sawtooth model, and on the signatures of the supersolid phase. Finally, we summarize our conclusions in Sec.~\ref{sec:Conclusions}.

\begin{figure}[t]
\begin{center}
\includegraphics[width=\columnwidth]{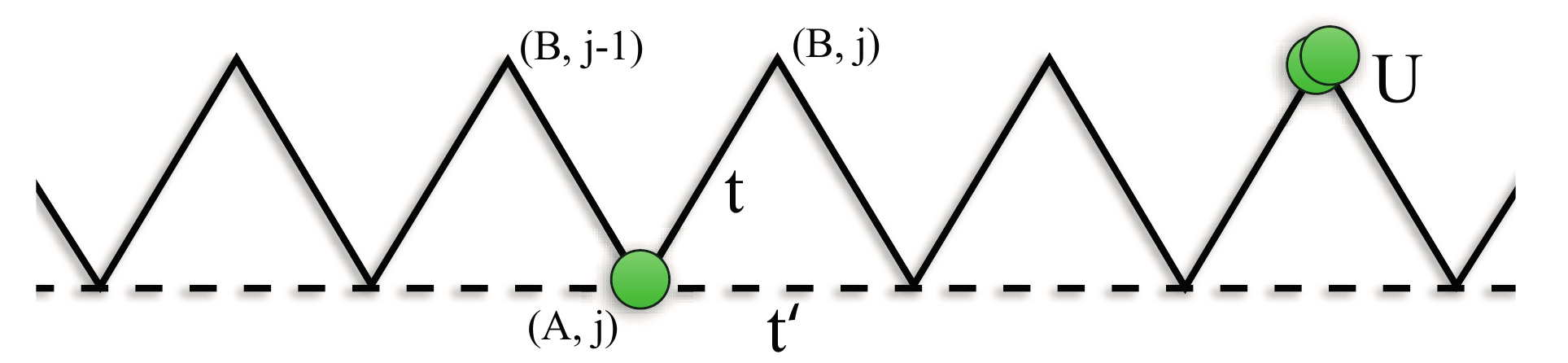}
 \end{center}
 \vspace*{-0.5cm}
\caption{(Color online) Sawtooth lattice discussed in the paper. The V-shaped plaquettes are un-frustrated, 
whereas the $\triangle$-shaped plaquettes are frustrated, since we assume $t>0$ and $t'<0$. The presence of frustrated and unfrustrated plaquettes is crucial for the 
super solid mechanism discussed in the paper.}
\label{fig:ladder}
\end{figure}



\section{Model}
\label{sec:Model}
We consider bosons in the sawtooth lattice as shown in Fig.~\ref{fig:ladder}(a), which is characterized by a hopping rate $t'<0$ along the lower leg, and 
a hopping rate $t>0$ along the rungs. The change in sign of $t'$, which may be achieved experimentally by e.g. lattice shaking~\cite{Struck2011}, introduces 
geometric frustration, a key ingredient below. The lower and upper legs have different coordination numbers, and hence 
constitute two distinct sublattices, which we denote as $A$ and $B$. For a sufficiently deep lattice, the system is described by the Bose-Hubbard model~(BHM):
\begin{align}
\mathcal{H} &= - t \sum_{i} (a_{i}^{\dagger}b_{i}^{\phantom \dagger} + b_{i}^{\dagger}a_{i+1}^{\phantom \dagger} + \text{H.c.})\nonumber\\
&+ t'\sum_{i}(a_{i}^{\dagger}a_{i+1}^{\phantom \dagger}+\text{H.c.}) 
 + \frac{U}{2}\sum_{\nu\in\{A,B\},i}  n^\nu_i(n^\nu_i-1)
\label{eq:ham}
\end{align}
where $a_i^{\dagger}$($b_i^{\dagger}$) and $a_i^{\phantom \dagger}$($b_i^{\phantom \dagger}$) are creation and annihilation operators
for bosons at site $i$ of leg A (B), and $n^A_i=a_i^{\dagger}a_i^{\phantom \dagger}$ ($n^B_i=b_i^{\dagger}b_i^{\phantom \dagger}$) is the number operator 
at site $i$ of leg A (B). Contact-like interactions lead to the on-site interaction term, characterized by the coupling constant $U$. Henceforth we consider all the 
physical quantities in units of $t=1$ which makes them dimensionless. 

\begin{figure}[t]
\begin{center}
\includegraphics[width=3in]{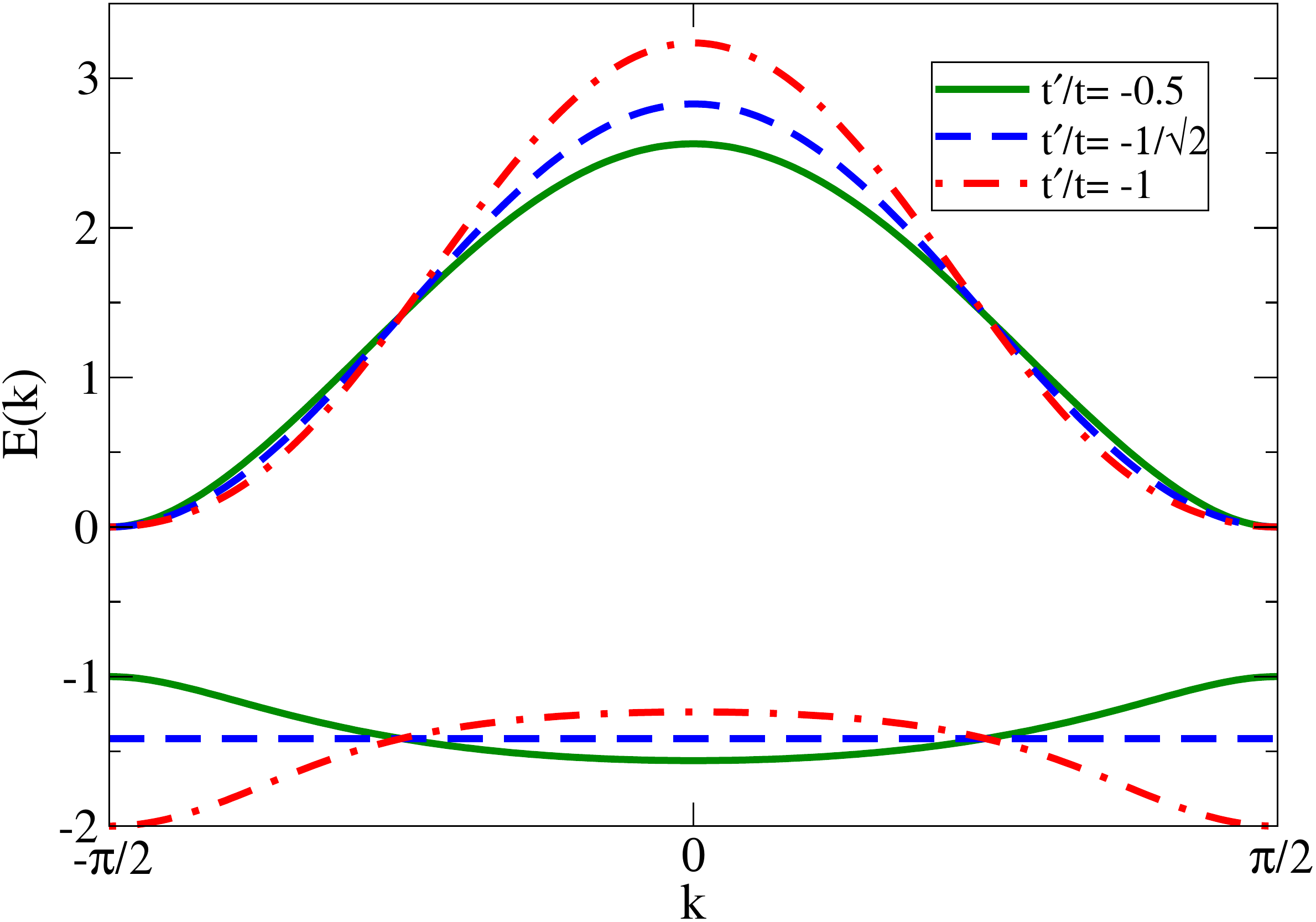}
\end{center}
\caption{(Color online) Single particle dispersion for Model (\ref{eq:ham}) with $U=0$ for different values of $|t'|/t$. }
\label{fig:flatband}
\end{figure}


\section{Non-interacting regime}
\label{sec:Non-interacting}

At first, it is convenient to study the non-interacting regime~($U=0$). In that case Model~\eqref{eq:ham} may be diagonalized in momentum~($k$) space 
(note that due to the broken sublattice symmetry, $k\in[-\pi/2,\pi/2]$):
$\hat H=\sum_k  [ E_\alpha(k) \alpha_k^\dag \alpha_k + E_\beta(k) \beta_k^\dag \beta_k  ]$, where the first~($\alpha$) and second~($\beta$) energy bands 
are characterized by the dispersion $E_{\alpha,\beta}(k)=-t'\cos 2k \mp 2t [ \cos^2 k +(t'/2t)^2 \cos^2 2k  ]^{1/2} $, and the 
bosonic operators $\alpha_k=\cos\theta_k a_k+\sin\theta_k b_k$,  and $\beta_k=-\sin\theta_k a_k+\cos\theta_k b_k$, with $\tan 2\theta_k=(2t/t') \cos k / \cos 2k$. 
The lowest band becomes flat at $|t'|=t/\sqrt{2}$ as shown in Fig.~\ref{fig:flatband}. 
When $|t'|< t/\sqrt{2}$ there is a single minimum in the lowest band, whereas two minima occur when $|t'|>t/\sqrt{2}$. 

As recently shown~\cite{Huber2010}, for the interacting case
 the flat band results in a solid phase at lattice filling $\rho=1/4$, but the solid order breaks upon doping due to the 
proliferation of domain walls. 
In addition, at $|t'|=t/\sqrt{2}$ the minimum of $E_\alpha(k)$ changes from $k=0$ to $k=\pi/2$. 
As a result, for non-interacting bosons, a transition occurs from a superfluid phase at $k=0$~(SF$_0$) 
to a superfluid phase at $k=\pi/2$~(SF$_{\pi/2}$).  Since $\alpha_{\pi/2}=a_{\pi/2}$, in the SF$_{\pi/2}$ the $B$ leg is depopulated.


\section{Ground-state phase diagram} 
\label{sec:Ground-state}

The ground state phase diagram is obtained by means of numerical density matrix renormalization group~(DMRG) 
calculations~\cite{White1992,Schollwock2005,footnote-even}. For our calculation we consider  up to $L=160$ sites, with maximally 
$4$ particles per site, and retaining up to $500$ states in the density matrix. 
In order to obtain the signature of possible quantum phase transitions in the system 
we compute the corresponding order parameters. Particularly relevant is  
the density structure factor 
\beq
S(k)=\frac{1}{L^2}\sum_{i,j}{e^{ik(i-j)}}\langle{n_{i}n_{j}}\rangle,
\eeq
where $\langle{n_{i}n_{j}}\rangle$ is the density-density correlation between sites $i$ and $j$.

In order to quantify the critical point for the SF$_0$ to Mott-insulator~(MI) transition discussed below, we calculate the 
Luttinger parameter $K$, which is the power-law exponent of the decay of the single particle 
correlation function $\Gamma_r =\langle a_i^\dagger a_j\rangle$. 
However, in order to obtain the value of $K$ we evaluate the structure factor $S(k)$ in the long wave-length limit~($k \rightarrow 0$), 
since in this limit~\cite{Giamarchi2004}: 
\beq
S(k) \propto K |k|/2\pi.
\label{eq:k}
\eeq
We also compute the single particle gap to characterize the gapped MI phase as 
\beq
E_{G}=\mu^+ -\mu^- ,
\label{eq:gap}
\eeq
where $\mu^+$~($\mu^-$) is the chemical potential for adding~(removing)  
one particle.
Finally, the momentum distribution function
\beq
N(k)=\frac{1}{L}\sum_{i,j}{e^{ik(i-j)}\langle{a^{\dagger}_{i}a_{j}}\rangle}.
\label{eq:mom}
\eeq
is particularly interesting as a possible experimental signature of the supersolid phase discussed below. 

In the following subsections we discuss the ground-state phase diagram at unit filling, and away from unit filling.

\subsection{Unit filling}

\begin{figure}[t]
\begin{center}
\includegraphics[width=3in]{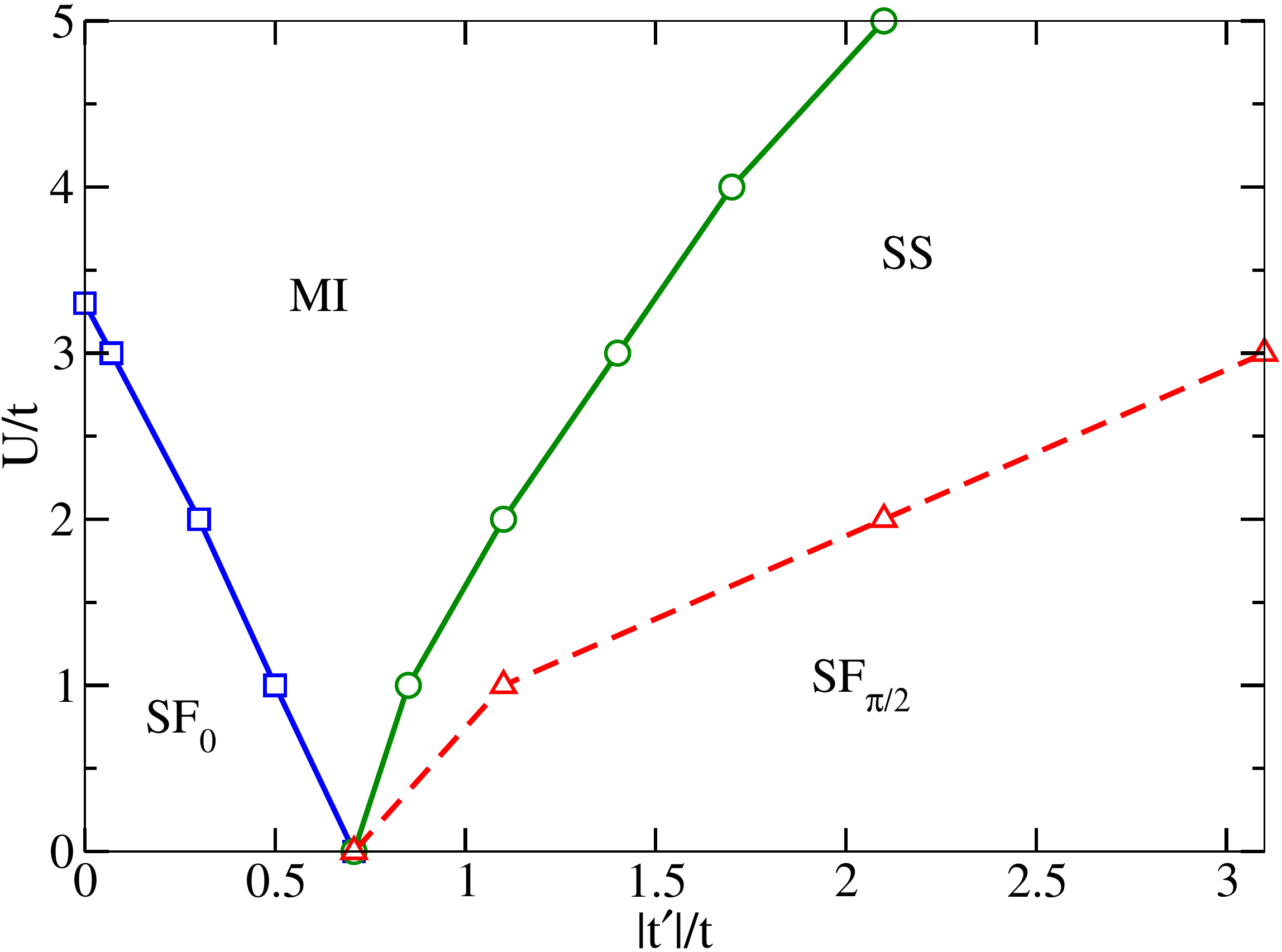}
 \end{center}
\caption{(Color online) Phase diagram of Model~(\ref{eq:ham}) for $\rho=1$ as a function of $|t'|/t$ and $U/t$~(see text).}
\label{fig:phasedia}
\end{figure}

Figure~\ref{fig:phasedia} shows the ground state phase diagram of model~(\ref{eq:ham}) at unit filling, $\rho=1$. 
As discussed above, in the non-interacting limit the system exhibits  
two gapless superfluid phases, SF$_0$ and SF$_{\pi/2}$. 
In addition to these phases, the system opens as expected a MI
phase at small $U$, which at the flat-band point is maintained all the way down to vanishing $U$. 
The second additional phase is, in contrast, rather unexpected. 
As we discussed below, this phase is a supersolid~(SS), 
which rather surprisingly occurs at $\rho=1$ and in the absence of inter-site interactions. 

\begin{figure}[b]
\begin{center}
\includegraphics[width=3in]{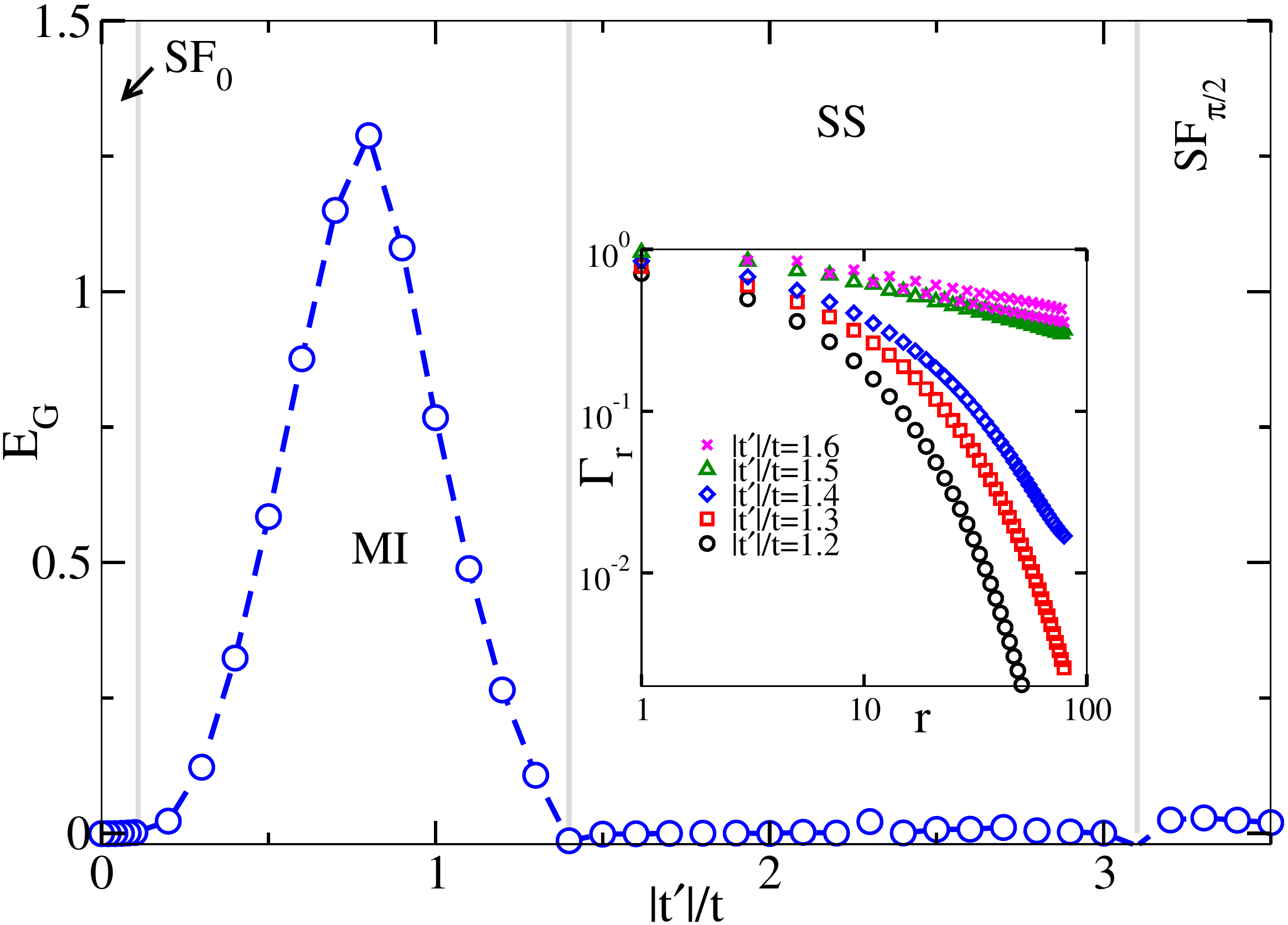}
\end{center}
\caption{(Color online) Extrapolated gap $E_{G}$ as a function of $|t'|/t$. The inset shows the exponential 
to power-law decay of the single particle correlation function $\Gamma_r$ at the MI-SS boundary. Only odd $r$ is shown to avoid oscillations.
\label{fig:gap}}
\end{figure}

\begin{figure}[t]
\begin{center}
\includegraphics[width=\columnwidth]{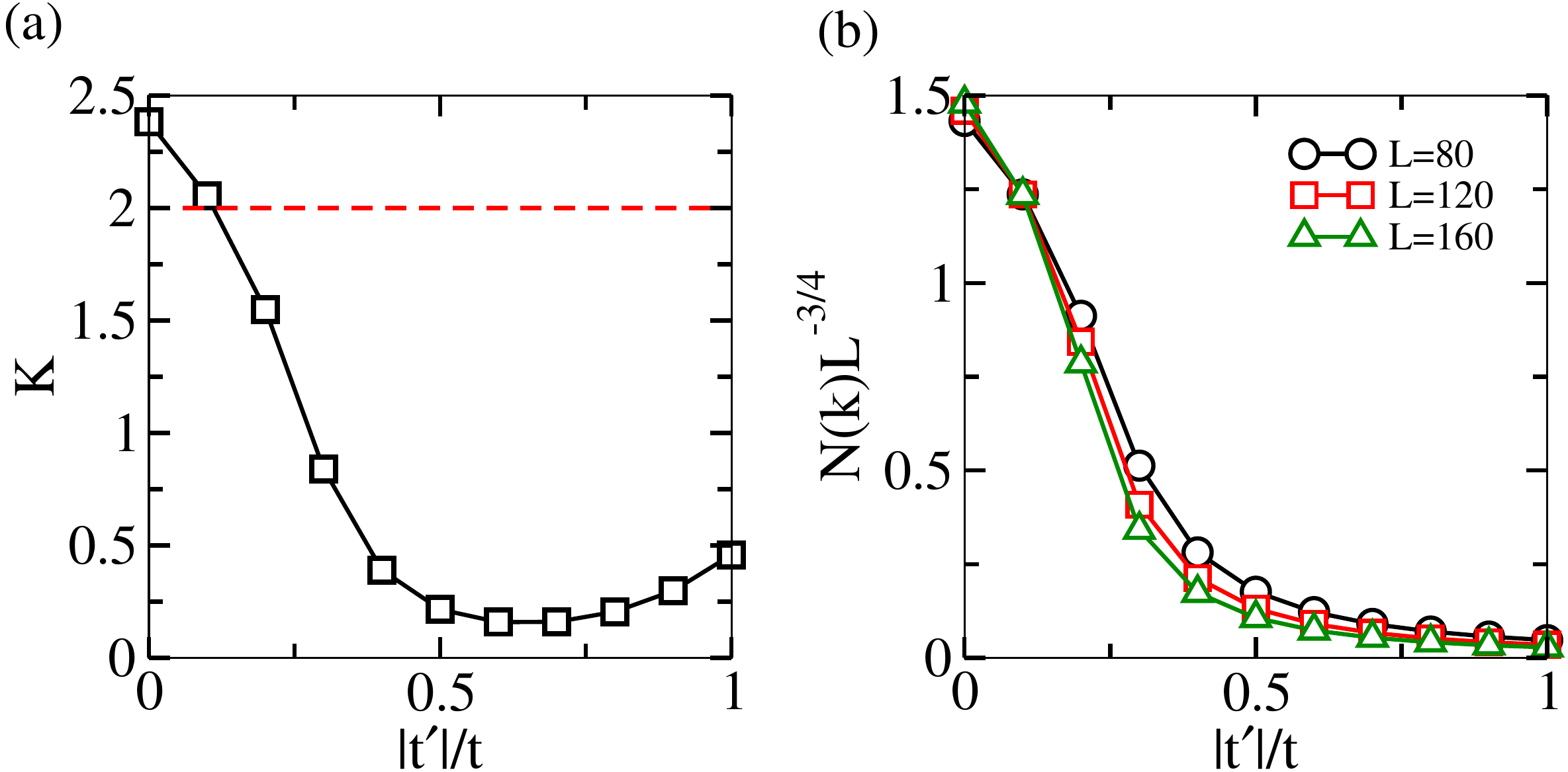}
\end{center}
\caption{(Color online) (a) Extrapolated values of K~(black squares) with respect to $|t'|/t$ for $U/t=3$. The $K=2$ line~(red dashed) is 
drawn to find the critical point for the SF$_0$-MI transition; (b) Scaled momentum distribution $N(k)L^{-3/4}$ for different lengths as a function of 
$|t'|/t$ for $U/t=3$.
\label{fig:knk}}
\end{figure}

\begin{figure}[b]
\begin{center}
\includegraphics[width=3in]{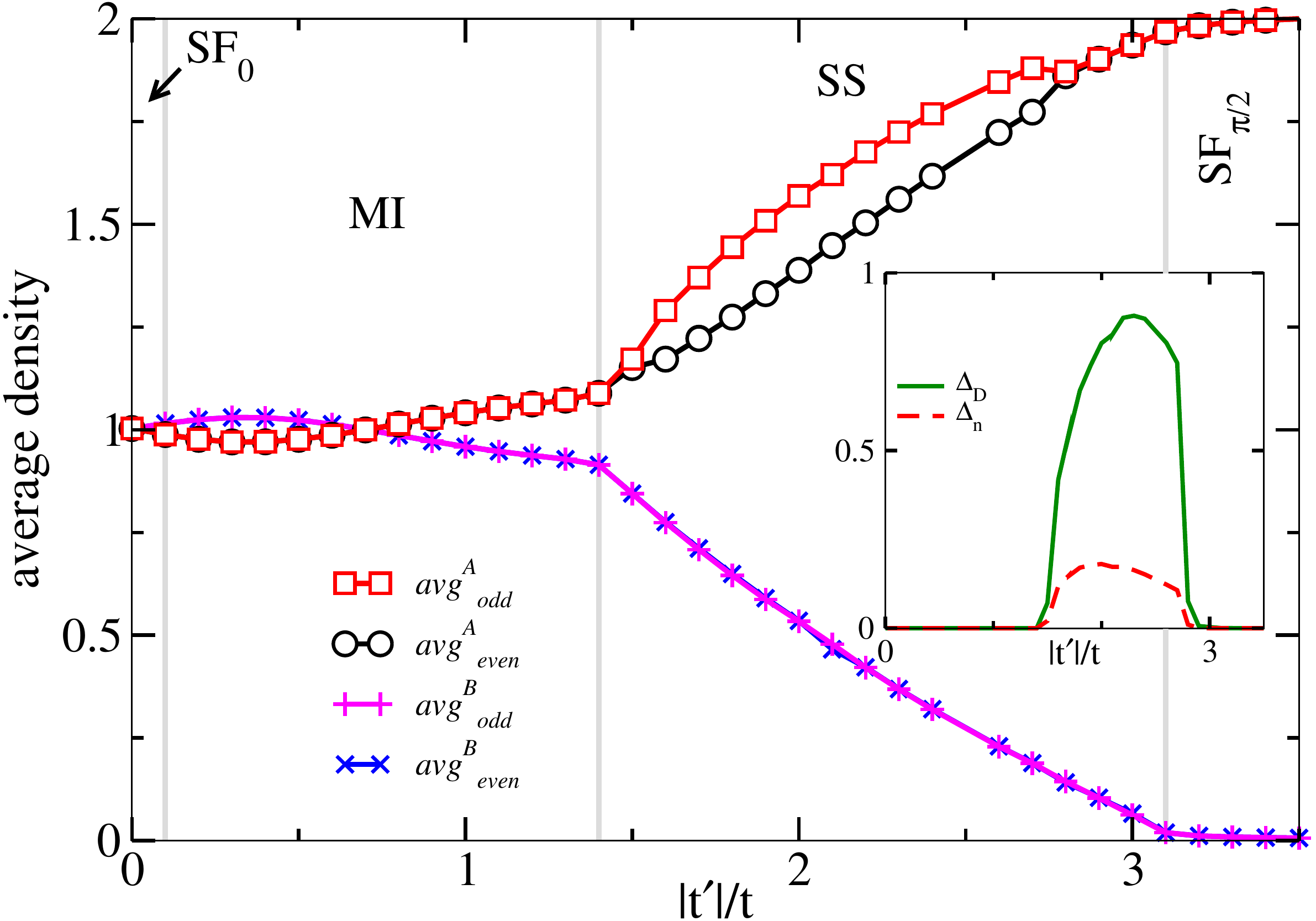}
\end{center}
\caption{(Color online) Number fluctuation between odd and even sites of the $A$ and $B$ legs, the inset shows $\Delta_n$ and $\Delta_D$.
\label{fig:delta}}
\end{figure}

\begin{figure}[t]
\begin{center}
 \includegraphics[height=2.in]{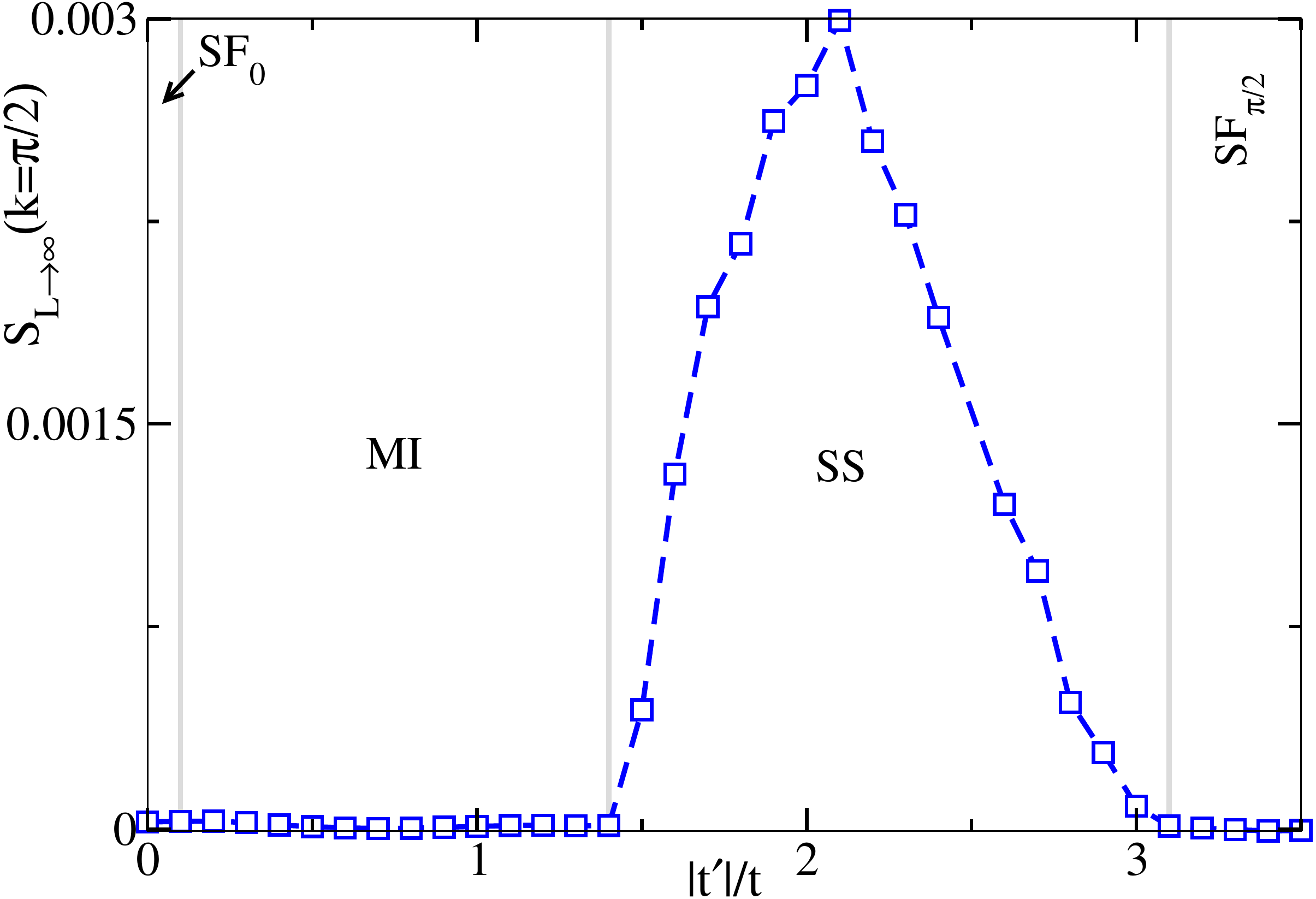}\hfill
 \includegraphics[height=2.05in]{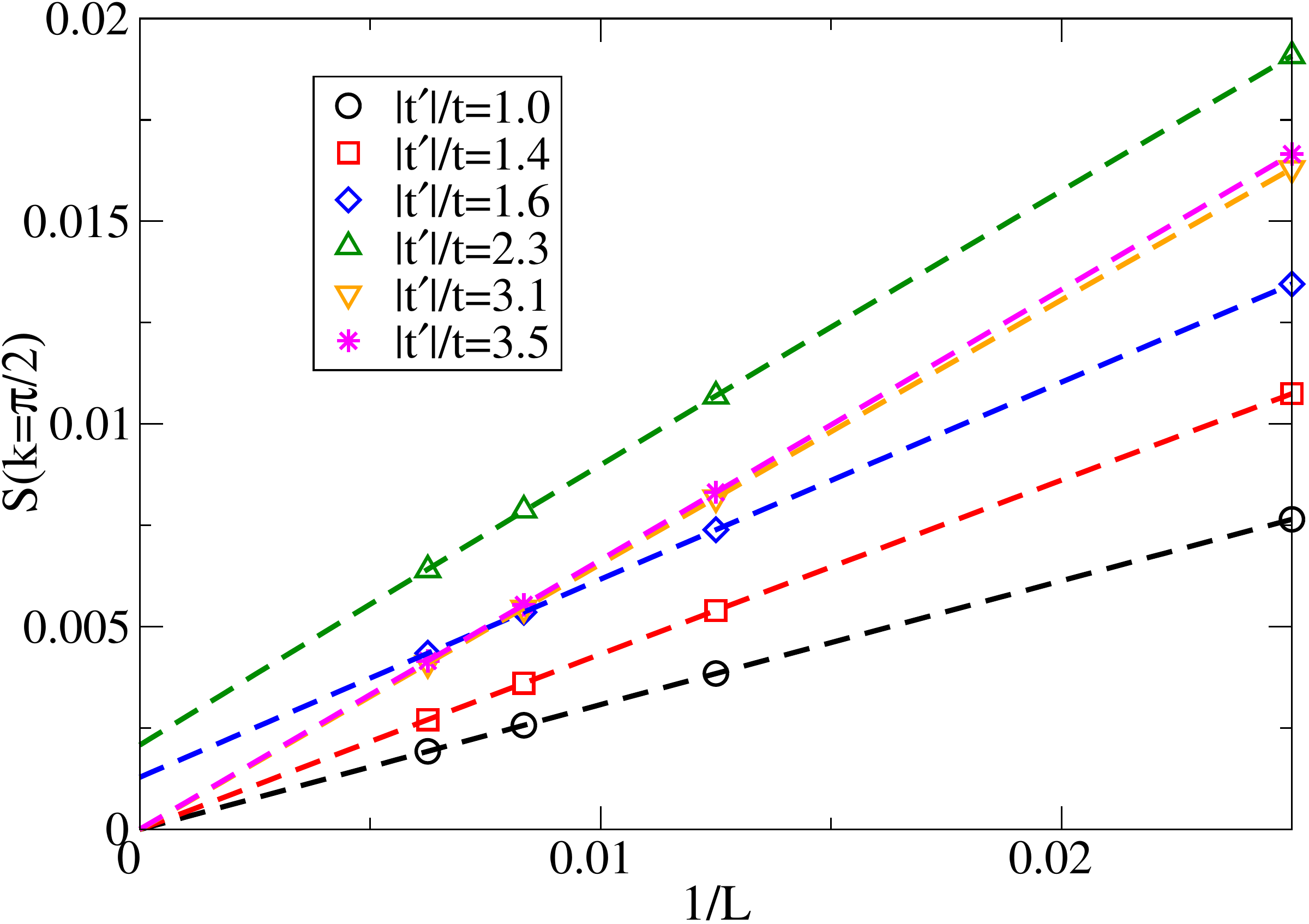}
\end{center}
\caption{(Color online) (Top) Scaled structure factor $S(k=\pi/2)$ with respect to $|t'|/t$ for $U/t=3.0$; 
(Bottom) Finite-size scaling of $S(k=\pi/2)$  for $U/t=3.0$ and different $|t'|/t$ values.
\label{fig:str}}
\end{figure}

In the following we discuss in detail the different phases for the case $U/t=3$.  
In Fig.~\ref{fig:gap} we plot the thermodynamic limit values of the single-particle gap~\eqref{eq:gap}
. The MI region is characterized by a finite gap, whereas $E_{G}\to 0$ in all other regions. 
The smooth gap opening at small $|t'|/t$ marks the Berezinskii-Kosterlitz-Thouless~(BKT) SF$_{0}$-to-MI 
transition, characterized by a Luttinger parameter $K=2$~\cite{Giamarchi2004}.  By performing a finite size scaling of $K$ computed using Eq.~\eqref{eq:k}, 
we obtain that the transition for SF$_0$ to MI phase occurs at $|t'|/t\simeq 0.1$~(see Fig.~\ref{fig:knk}(a)). 
The BKT transition is further confirmed by the scaling of the momentum distribution, $N(k=0)\propto L^{1-\frac{1}{2K}}$~\cite{dhar1,dhar2,Greschner2013}. 
In Fig.~\ref{fig:knk}(b) we show that the values of  
$N(k=0)L^{-3/4}$ for different lengths (L=80, 120, 160) intersect at the transition point $|t'|/t\simeq 0.1$.

In contrast, the sudden vanishing of $E_{G}$ for larger $|t'|/t$~(at $|t'|/t= 1.4$ in Fig.~\ref{fig:gap}) 
does not match with a BKT transition. 
The gapped to gapless transition is, however, confirmed by the  
single-particle correlation function $\Gamma_r=\langle a_i^\dag a_{i+r} \rangle$, 
which decays exponentially in the MI and algebraically for $|t'|/t>1.4$~(inset of Fig.~\ref{fig:gap}). This 
transition is likely to be a weak first order transition which could not be confirmed in our numerics. 

The superfluid region opening at the large $|t'|/t$ side of the MI turns out to be a SS phase.
Figure~\ref{fig:delta} shows the 
average density $avg^A_{odd,even}$~($avg^B_{odd,even}$) at even and odd sites in the A~(B) leg. 
The difference of densities at both legs is not surprising due to the asymmetry between the legs. 
More interesting is the behavior within a given leg. Whereas for $avg^B_{odd}=avg^B_{even}$ for all $|t'|/t$, 
$avg^A_{odd}\neq avg^A_{even}$ in the SS region~($1.4 < |t'|/t < 3.1$ in Fig.~\ref{fig:delta}) indicating a clear density modulation in this region. 
The spontaneously broken translational symmetry along the $A$ leg is characterized by the amplitude $\Delta_n=avg^A_{odd}-avg^A_{even}$ of the 
odd-even modulation~(inset of Fig.~\ref{fig:delta}). The broken symmetry translates as well in a difference of bond-kinetic energy $\Delta_D=avg^D_{odd}-avg^D_{even}$, 
where $avg_{odd,even}^D=\frac{1}{L}\sum_{i \in A_{odd,even}}(b^\dagger _{i-1}a_i+a^\dagger _i b_{i+1}+H.c.)$, which is finite in the SS region~(inset of Fig.~\ref{fig:delta}). 
The finite $\Delta_D$ relates with the V-type dimerization discussed below.
The end of the SS region and the onset of the SF$_{\pi/2}$  phase is marked by the vanishing of both $\Delta_{n}$ and $\Delta_D$, and the depopulation of the $B$ leg. 

The SS phase can be further confirmed by a finite peak in $S(k)$ at non-zero wave vector $k$. 
However, we note that the sawtooth lattice breaks a trivial translational symmetry due to which $S(k)$ shows a peak at $k=\pm \pi$. 
Therefore, the presence of the SS phase in the system is revealed by a finite peak in the structure factor at $k=\pm \pi/2$, 
confirmed by extrapolation to the thermodynamic limit. In Fig.~\ref{fig:str}~(top) we show, for $U/t=3$, the 
extrapolated value of $S(k=\pi/2)$, which is finite in the region $1.4 < |t'|/t < 3.1$, i.e. the SS phase. The extrapolation of $S(k=\pi/2)$
is shown in detail in Fig.~\ref{fig:str}~(bottom).

\subsection{Away from unit filling}
At this point we consider incommensurate filling. This discussion is particularly relevant, since the presence of an overall harmonic trap, typical of experiments on atoms 
in optical lattices, results in an inhomogeneous density distribution. In Fig.~\ref{fig:phasedia2} we depict our results for $U/t=2$. 
In addition to the gapped density-wave phase at $\rho=0.25$ at the flat-band point~\cite{Huber2010}, we obtain 
gapped phases at $\rho=n/4$, for all integer $n\geq 1$ (black lines and squares in Fig.~\ref{fig:phasedia2}). 
The gapped phases at these densities are expected to occur at the large interaction
limit. The phases at $\rho=(2n+1)/4$ with $n>1$ are similar to the phase at $\rho=1/4$, being comparatively narrow due to small interaction.
On the contrary, the gapped phase at $\rho=1/2$ is a trivial insulator due to a filled band.
Even more relevant is the robustness of the SS phase, which extends over a large parameter region, for densities 
$\rho>0.5$. 


We illustrate the behavior of the system for the case of two representative values,
$|t'|=1/\sqrt 2$ and $1.5$~(brown dashed lines in Fig.~\ref{fig:phasedia2}). 

\bfig[t]
  \centering
  \includegraphics[width=3in]{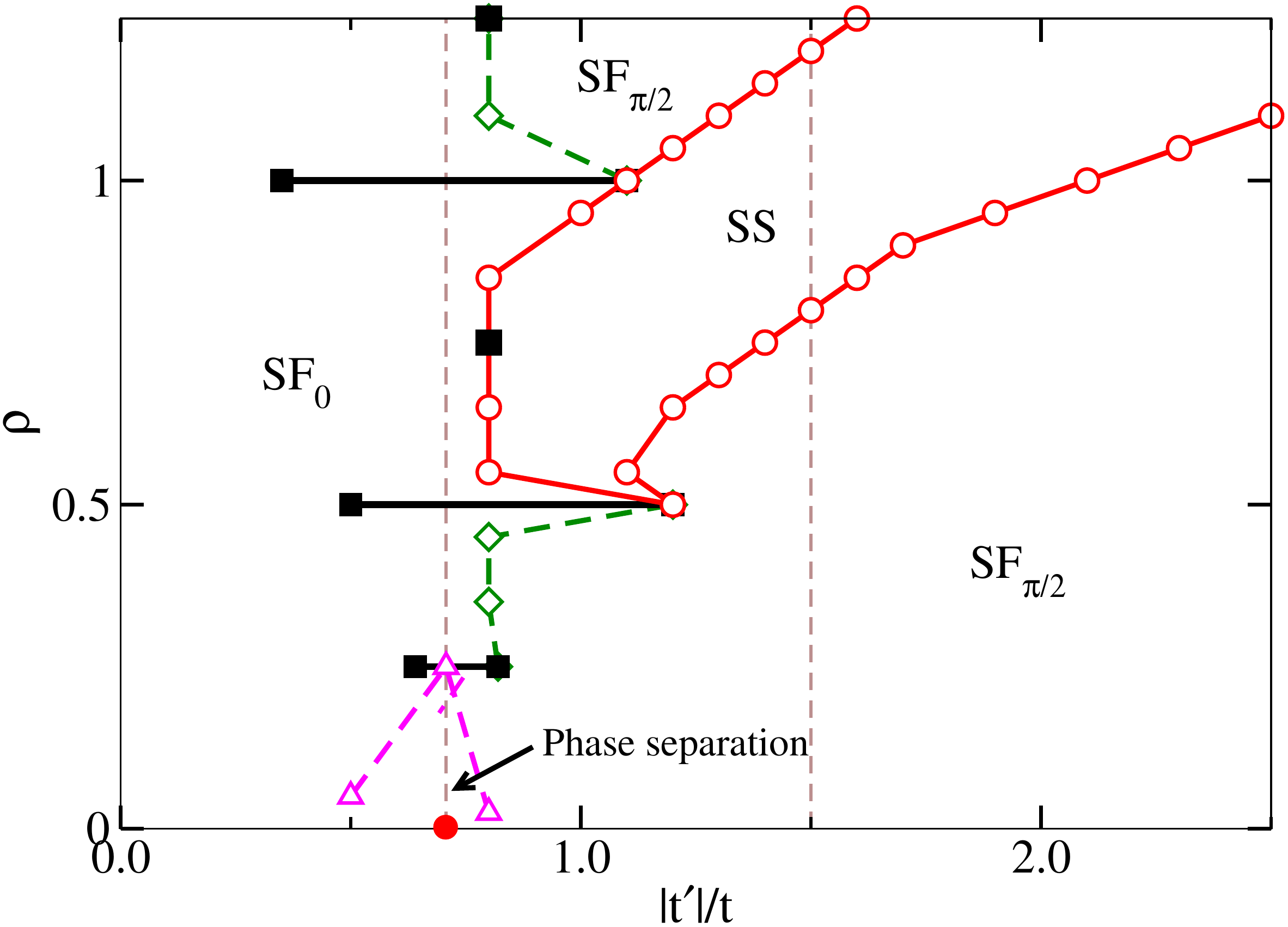}
    \caption{(Color online) Phase diagram of the sawtooth chain as a function of $\rho$ and $|t'|/t$ for $U/t=2.0$. 
    Black lines with squares show the gapped regions. The SS phase is 
    bounded by the red circles. The SF$_{\pi/2}$ phase occupies a large portion of the
    phase diagram below the SS region and for $|t'|/t > 1/\sqrt2$, also around $\rho=1$. The region bounded by 
    magenta triangles shows a macroscopic jump in the density. The brown dashed lines are two representative cuts at
$|t'|/t = 1/\sqrt2$ and $1.5$~(see text). The red dot corresponds to the flat-band point where $|t'|/t = 1/\sqrt2$ 
     }
    \label{fig:phasedia2}
\efig

\bfig[t]
  \centering
  \includegraphics[width=3in]{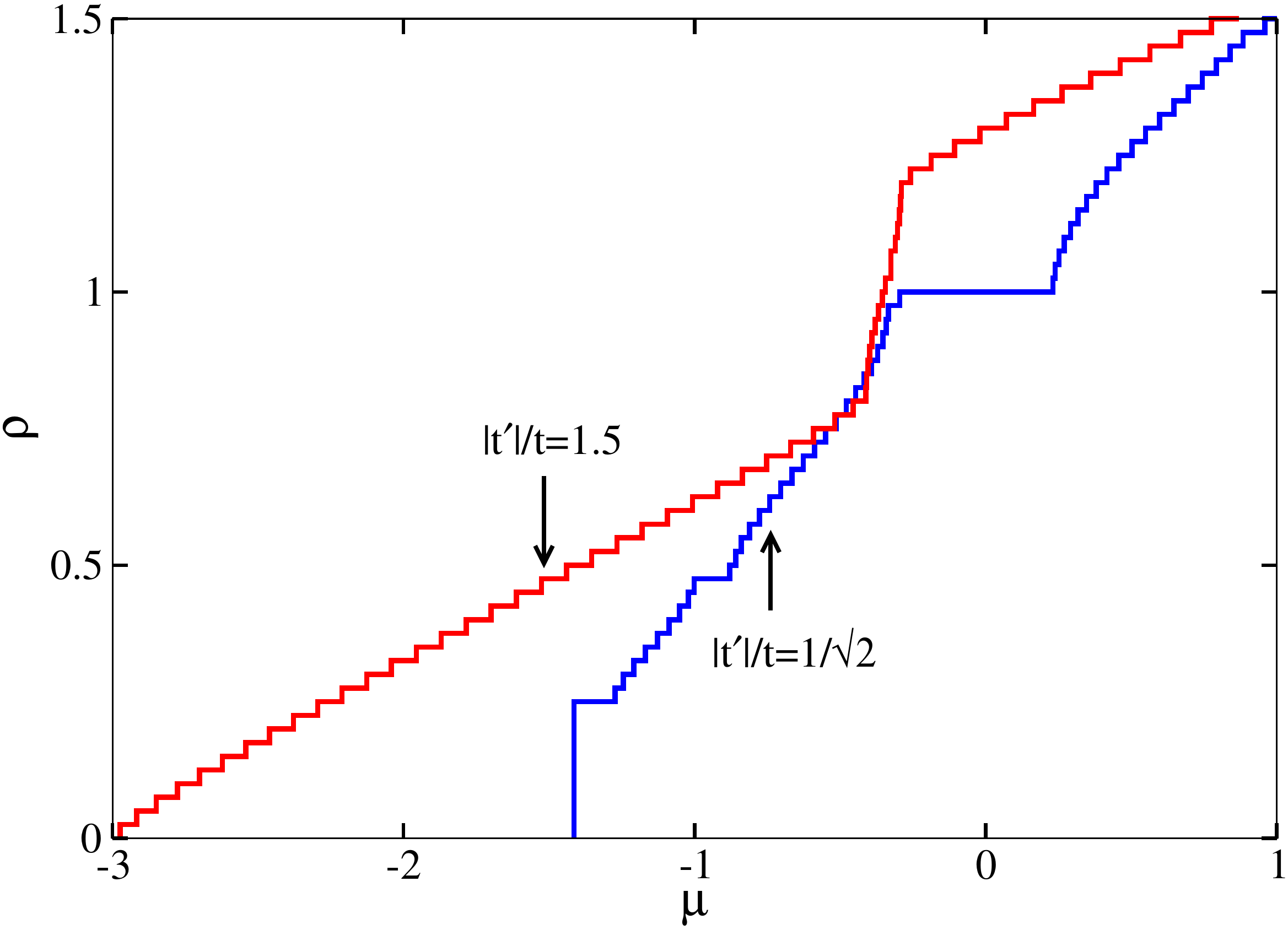}
    \caption{(Color online) $\rho(\mu)$ for $L=40$, $U/t=2$, and $|t'|/t=1/\sqrt 2$~(blue curve) and 
    $1.5$~(red curve).
     }
    \label{fig:rhomu}
\efig

In Fig.~\ref{fig:rhomu},
 we plot $\rho$ as a function of the chemical potential $\mu$ for $|t'|/t=1/\sqrt 2$. Due to band flatness at that point, 
at a critical $\mu$ there exists a macroscopic jump in the density up to $\rho=0.25$. This macroscopic jump  
exists for a small region around  $|t'|/t=1/\sqrt 2$. This region is marked in the phase diagram by magenta triangles.
At $\rho=0.25$ a gap appears marked by a plateau in $\rho(\mu)$. Then, the system becomes a gapless and compressible superfluid. 
Further increase in $\rho$ induces additional gaps as discussed above. 
In between the gapped plateaus the system is a superfluid. 

\bfig[b]
  \centering
  \includegraphics[width=3in]{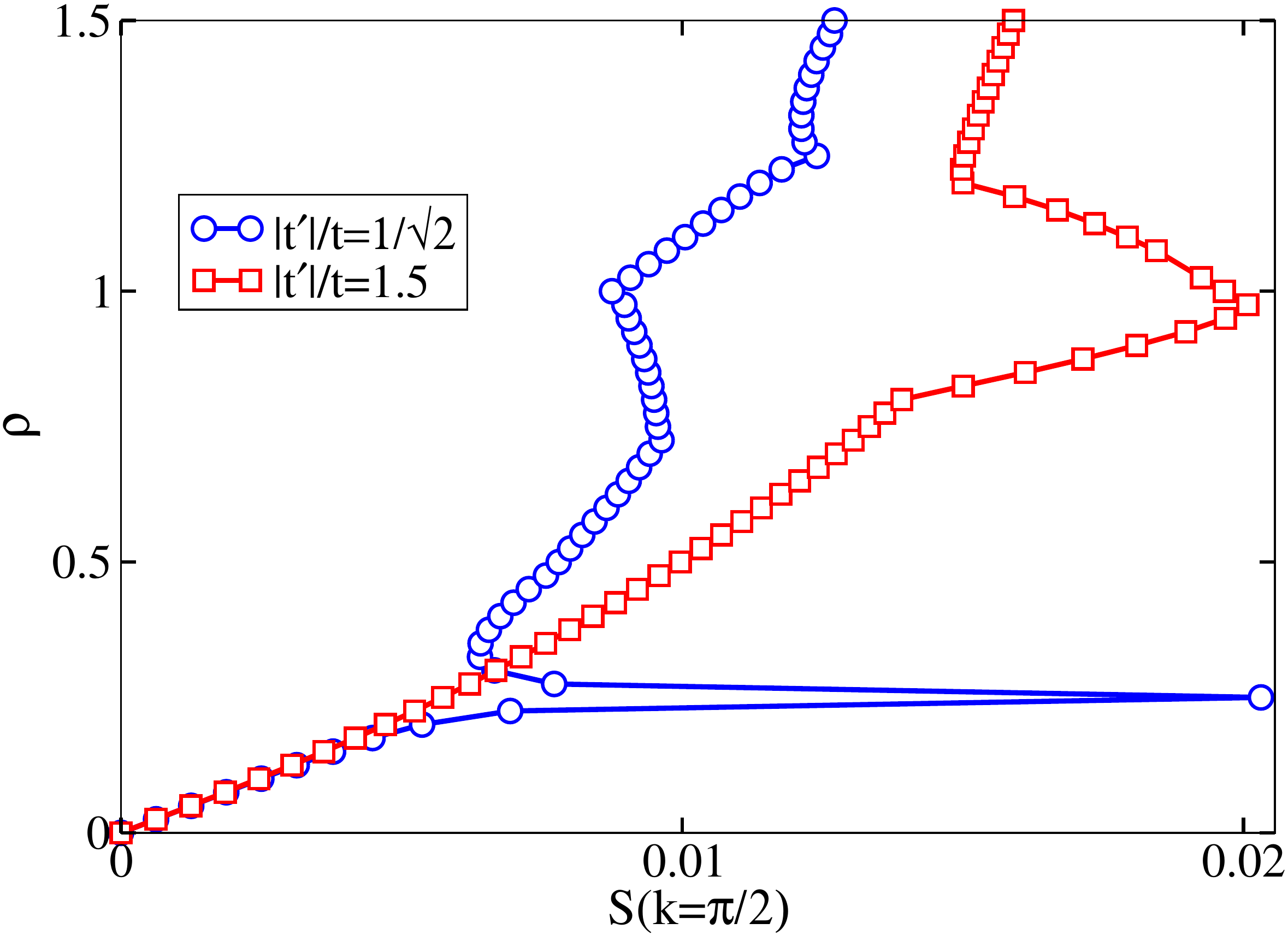}
    \caption{(Color online) Structure factor $S(k=\pi/2)$ for 
different fillings $\rho$ for $L=40$, $U/t=2$, and $|t'|/t=1/\sqrt 2$~(blue circles) and        
    $1.5$~(red squares).
     }
    \label{fig:strincomm}
\efig

The dependence of $\rho(\mu)$ is very different for $|t'|/t=1.5$, since for that value the band is not flat.
As a result the system starts at $\rho\to 0$ in the SF$_{\pi/2}$ phase. 
In Fig.~\ref{fig:rhomu} we observe that the curve $\rho(\mu)$ is continuous 
with two kinks at $\rho\simeq 0.8$ and $1.2$.
These kinks correspond to the transition to the SS phase. The abrupt growth of $\rho$ in the SS phase, shows that the SS phase 
is highly compressible. 

Finally, Fig.~\ref{fig:strincomm} shows 
$S(k=\pi/2)$  for $|t'|/t=1/\sqrt 2$ and $1.5$. For $|t'|/t=1.5$, 
$S(k=\pi/2)$ increases in the SS phase, $0.8<\rho<1.2$. In contrast, for $|t'|/t=1/\sqrt 2$,
$S(k=\pi/2)$ remains small except at three peaks at $\rho=0.25, 0.75 $ and $1.25$.
The two small peaks at $\rho=0.75 $ and $1.25$ are due to the presence of two small density-wave phases shown as 
black squares in Fig.~\ref{fig:phasedia2}. 
The sharp peak at $\rho=0.25$ corresponds to the density-wave phase reported in Ref.~\cite{Huber2010}.

\section{Weak coupling limit} 
\label{sec:Weak-coupling}
\begin{figure}[b]
\begin{center}
\includegraphics[width=\columnwidth]{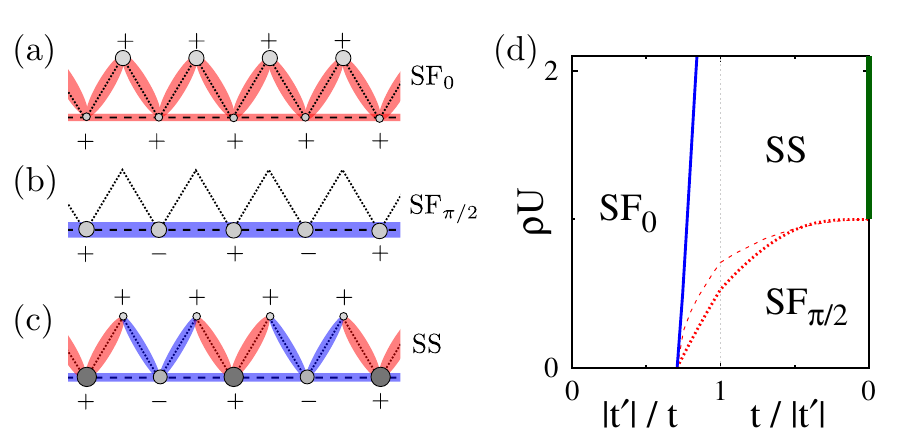}
\end{center}
\vspace{-0.6cm}
\caption{(Color online) Sketch of the SF$_0$~(a),  SF$_{\pi/2}$~(b) and SS~(c). 
The circle size is proportional to the local density,  whereas $\pm$ denote the sign.
(d) Phase diagram in the weak-coupling limit as function $U \rho$ and $|t'|/t$ for a fixed~(but arbitrary) filling. 
The SF$_0$ to the SS phase is shown as a solid line. The dotted and dashed lines depict, respectively, the SF$_\pi/2$-SS 
transition line obtained 
from the variational approach~\eqref{eq:cs} and the roton instability. 
At $t=0$ SS-order vanishes (thick solid line) resulting in an highly degenerate ground-state, 
since particles may occupy the uncoupled $B$ 
sites in an arbitrary configuration.}
\label{fig:cs}
\end{figure}
An intuitive insight on the emergence of the SS is
obtained from the classical limit of Model~\eqref{eq:ham}. We assume each site to be in a coherent state with a well defined density and phase, $0$ or $\pi$, 
corresponding to the two possible minima of $E_\alpha(k)$.
We consider a simplified model in which $b_j=\eta$, $a_{2j}=\xi$, and $a_{2j+1}=\chi$, such that we allow for both a possible density imbalance 
between the $A$ and $B$ legs, and for an even-odd asymmetry in the $A$ leg. We may hence minimize the energy, which 
without loss of generality may be calculated for a four-site unit cell:
\begin{align}
\langle \mathcal{H} \rangle &= - 4 t (\xi \chi + \chi \eta) - 4 t' \xi \eta +\nonumber\\
&+ U \left( \xi^4 + 2\chi^4 + \eta^4 \right)- \mu \left( \xi^2 + 2\chi^2 + \eta^2 \right)
\label{eq:cs}
\end{align}
Within this approach the phase diagram splits into three regions~(Fig.~\ref{fig:cs}(d)). 
For $|t'|/t < 1/\sqrt{2}$ the three coefficients have the same sign, and the particles occupy both $A$ and $B$ sites 
corresponding to the SF$_0$-phase~(Fig.~\ref{fig:cs}(a)). For small $U/t$ and $|t'|/t >1/\sqrt{2}$, the $B$ sites depopulate~($\eta=0$), the density is homogeneous in the $A$ sites~($|\chi|=|\xi|$), and 
${\mathrm sign}(\chi)\neq {\mathrm sign}(\xi)$, corresponding to the SF$_{\pi/2}$ phase~(Fig.~\ref{fig:cs}(b)).  
A sufficiently strong repulsive interaction $U/t>(U/t)_c$ redistributes population to the $B$ sites.
However, how the particles re-distribute in the $A$ leg is crucially determined by the existence of frustrated and un-frustrated plaquettes in the sawtooth lattice.
 In order to minimize kinetic energy particles favor the un-frustrated V-shaped plaquettes of the sawtooth forming V-shaped dimers. 
As a result particles break the translational symmetry spontaneously, preferably occupying every second V-plaquette~(Fig.~\ref{fig:cs}~(c)), which leads to a density 
modulation in the A sites, that characterizes, as mentioned above, the SS phase. Note that this simple picture also predicts a finite $\Delta_D$ in the SS region, as observed in the numerics.

\begin{figure}[t]
\begin{center}
\includegraphics[width=0.46\linewidth]{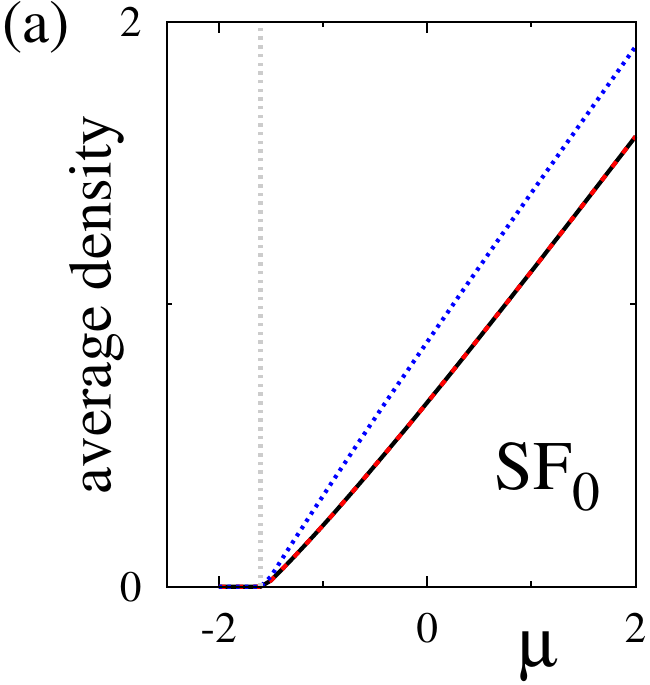}
\includegraphics[width=0.46\linewidth]{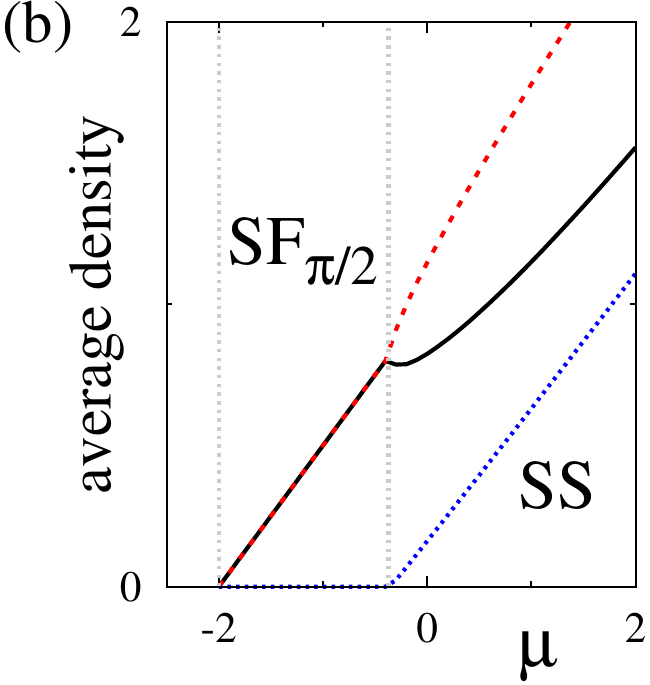}
\end{center}
\caption{(Color online)Densities of the classical model for $\rho U=2$ and (a) $t=1$ and $t'=-0.5$ and (b) $t=0.8$ and $t'=-1$ as a function
 of the chemical potential $\mu$. The dotted (blue) line depicts the average density of the B sites, 
the solid (black) and dashed (red) lines show densities of odd and even A sites. \label{fig:cscut}}
\end{figure}

As discussed above, already the classical or weak -coupling limit gives an intuitive insight into the formation of the SS phase. 
We provide in this section some additional details. 
In Fig.~\ref{fig:cscut} we show typical densities as function of $\mu$, obtain using Model~(\ref{eq:cs}). 
While in the SF$_0$ and SF$_{\pi/2}$ phases the A-sites do not exhibit any density modulation, when entering the 
SS-phase the $avg^A_{even}$ and $avg^A_{odd}$ curves separate from each other. 
The amplitude of the modulation on the $A$ sites in the limit $\mu\gg t,|t'|,U$ is given by $\xi-\chi = \frac{2 t}{\sqrt{U \mu}}$, 
is suppressed with increasing repulsion $U$. Within this approach the population of the 
$B$ sites in the SF$_{\pi/2}$ phase is strictly zero. 

We determine the SF$_0$ to SS transition as first-order, whereas the SF$_{\pi/2}$ to SS 
transition is a second-order phase transition exhibiting a discontinuity in in second derivative of the ground-state energy.



\section{Roton instability.} 
\label{sec:Roton}

Insight on the nature of the SF$_{\pi/2}$-SS transition is obtained in the limit in which $U\ll E_\beta(k)-E_\alpha(k)$ for all $k$. 
In that case we may project Model~\eqref{eq:ham} onto the lowest energy band:
\begin{equation}
\!\! H\!\simeq \!\! \sum_k \! E_\alpha(k)\alpha_k^\dag\alpha_k\! +\! \frac{U}{2}\!\!\sum_{q,k,k'}\!\! f_{k,k'}^{k+q,k'-q} \alpha_{k+q}^\dag \alpha_{k'-q}^\dag \alpha_{k'} \alpha_k,
\label{eq:hamk}
\end{equation} 
with $f_{k_1,k_2}^{k_3,k_4}=\prod_{l=1}^4 \cos(\theta_{k_l})+\prod_{l=1}^4 \sin(\theta_{k_l})$. Note that, although the on-site interactions are contact-like, 
the effective interactions are momentum dependent. 

Starting from Eq.~\eqref{eq:hamk} and
after expanding the interaction part up to second order terms assuming condensation at $q=\pi/2$, i.e. $\alpha_q \simeq \sqrt{N} + \tilde{\alpha}_q$ 
the total Hamiltonian may be written up to constants as
$\mathcal{H} \simeq \sum_{k>0} A(k) \left(\tilde{\alpha}_k^\dagger \tilde{\alpha}_k + \tilde{\alpha}_{-k}^\dagger \tilde{\alpha}_{-k} \right)
+ B(k) \left(\tilde{\alpha}_k^\dagger \tilde{\alpha}_{-k}^\dagger + \tilde{\alpha}_{k} \tilde{\alpha}_{-k} \right)$
with 
\begin{align}
A(k) &= E_\alpha(k) + \rho U \left[ \cos 2\theta_q ( \cos2\theta_k - \cos2\theta_q ) \right. \nonumber\\
&+ \left. ( 1+\cos2\theta_q \cos2\theta_k) \right]\nonumber\\
B(k) &= \rho U \left( 1+\cos2\theta_q \cos2\theta_k \right)
\label{eq:hambogoliubov}
\end{align}
This Hamiltonian may be readily diagonalized using a Bogoliubov transformation
$\beta_k = \cosh \gamma_k \tilde{\alpha}_k - \sinh \gamma_k \tilde{\alpha}_{-k}^\dagger$ yielding the Bogoliubov spectrum of excitations,
$\epsilon(k)^2=\left(\widetilde{E_A}(k) + 2 U \rho (1 + \cos 2\theta_k) \right) \widetilde{E_A}(k)$ 
with $\widetilde{E_A}(k)=E_A(k) -2t' + U \rho (\cos 2\theta_k -1)$.

\begin{figure}[t]
\begin{center}
\includegraphics[width=\columnwidth]{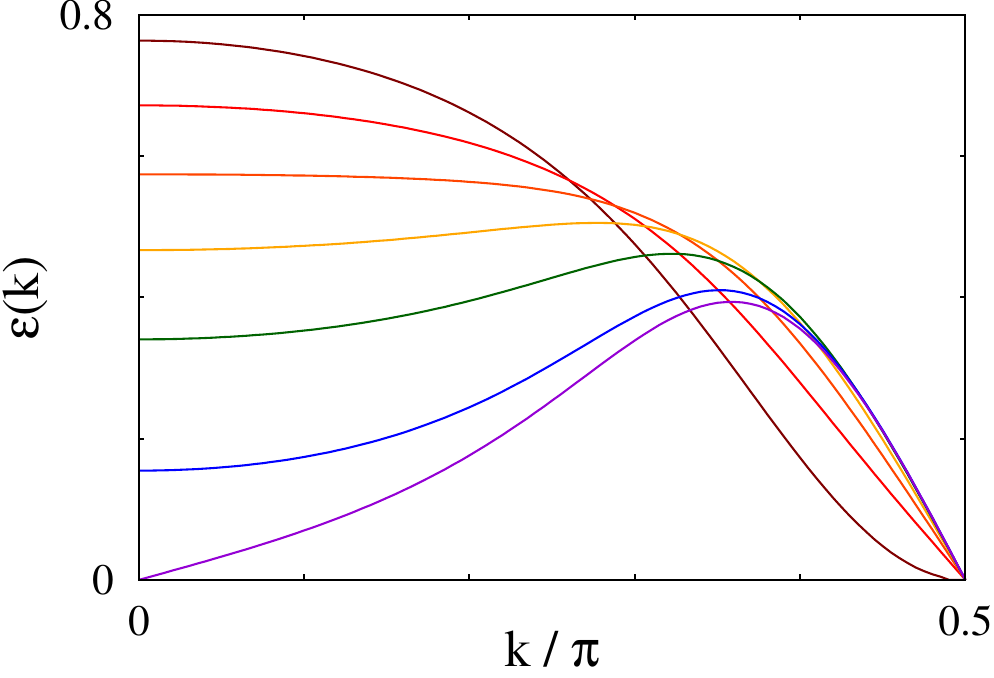}
\end{center}
\caption{(Color online)Bogoliubov spectrum $\epsilon(k)$ for $|t'|/t=1$ and (from top to bottom) 
$\rho U=0$, $0.1$, $0.2$, $0.3$, $0.4$, $0.5$, $0.5278$. At $k=0$ a roton-like minimum develops, which touches zero at  $\rho U \simeq 0.5278$. \label{fig:bogroton}}
\end{figure}

In Fig.~\ref{fig:bogroton} we depict the corresponding energy spectrum emerging for finite $\rho U$.
As usual the spectrum exhibits a linear (phonon-like) 
dispersion for $k$ close to $\pi/2$. Interestingly, for finite $\rho U$ it acquires a local minimum at $k=0$, that resembles  
the roton dispersion minimum of superfluid He~\cite{Landau1949}, and that occurs, as for dipolar condensates~\cite{Santos2003}, due to the momentum 
dependence of the interactions. For a critical value of $\rho U$ the roton-like minimum reaches zero energy, becoming unstable, marking the transition to the SS. 
As shown in Fig.~\ref{fig:cs}~(d) the critical $\rho U$ for roton instability agrees well with the SF$_{\pi/2}$-SS transition 
line obtained by the classical model. Hence, we can conclude that the SF$_{\pi/2}$ is destabilized through roton instability that leads to the SS phase.

\begin{figure}[t]
\begin{center}
\includegraphics[width=\columnwidth]{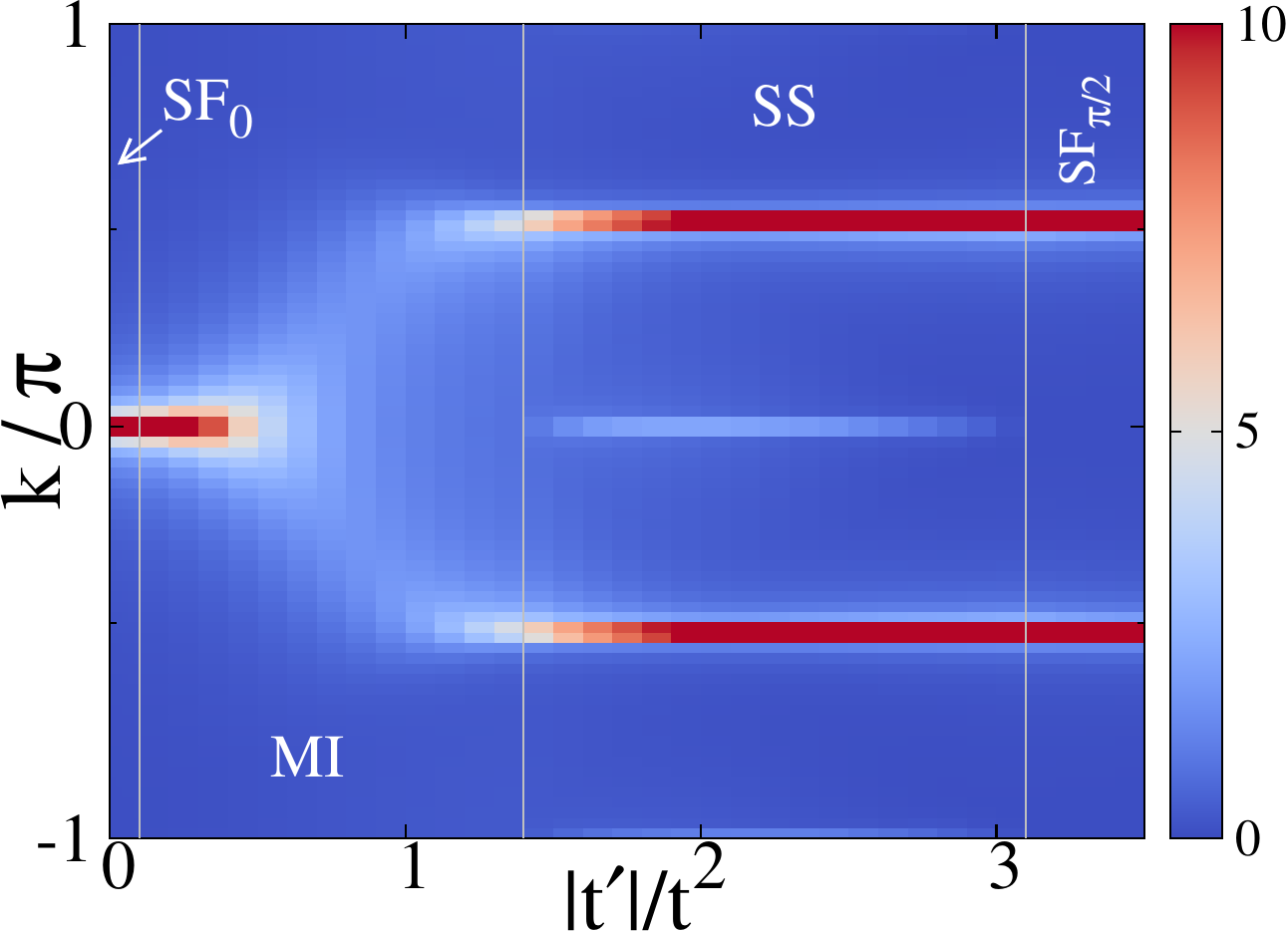}
\end{center}
\caption{(Color online)Momentum distribution $N(k)$ in different regions of the phase diagram shown in Fig.~\ref{fig:phasedia} for $U=3$.
\label{fig:md}}
\end{figure}

\begin{figure}[b]
\begin{center}
\includegraphics[width=3in]{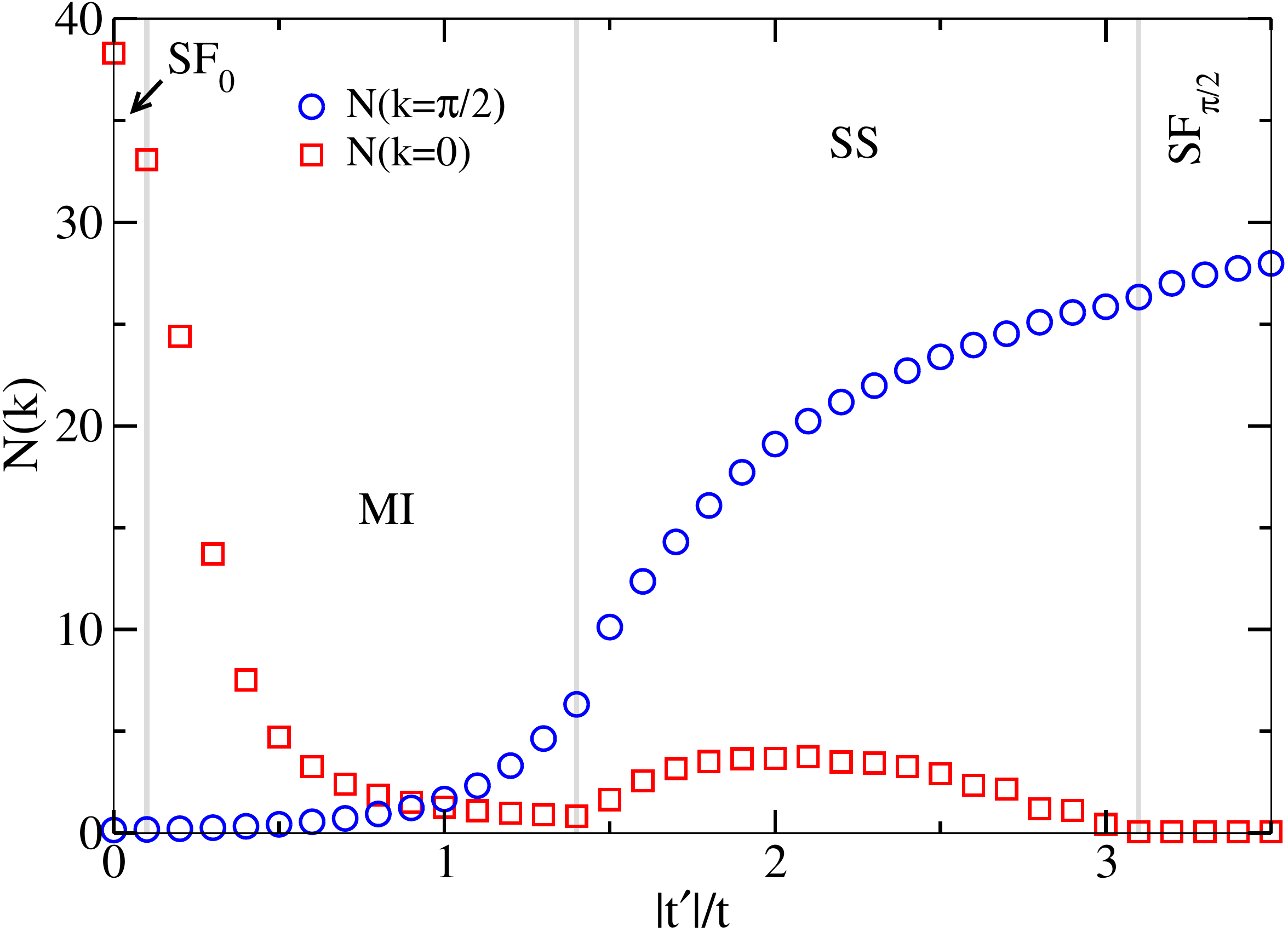}
\end{center}
\caption{(Color online) $N(0)~$(red squares) and $N(\pi/2)$~(blue circles) in different regions of the phase diagram shown in Fig.~\ref{fig:phasedia} at $U=3$.
\label{fig:md2}}
\end{figure}

\begin{figure}[t]
\begin{center}
\includegraphics[width=\linewidth]{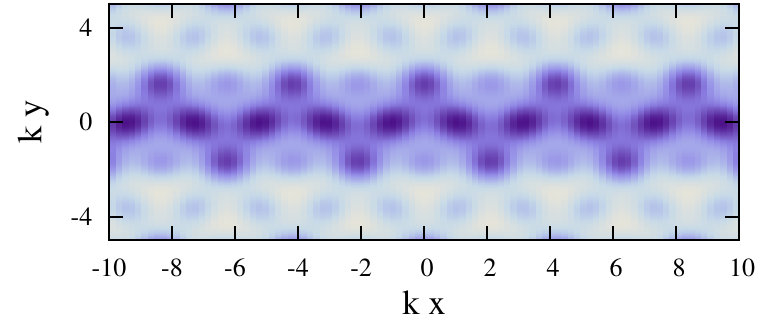}
\end{center}
\caption{(Color online) Sawtooth lattice formed by an incoherent superposition between a Kagome lattice~\cite{Jo2012}
 and an additional lattice $V({\vec r})=\sin^2\left(\frac{\sqrt{3}}{4} k y\right) + \frac{1}{4} \sin^2\left( \frac{\sqrt{3}}{2}  k y\right) $ with $k$ the laser wavenumber. Darker regions mean lower potential.
\label{fig:saw}}
\end{figure}

\section{Experimental signature}
\label{sec:Experimental}

In the section we briefly discuss about the possible signatures of the SS phase in optical lattice experients using ultracold atoms. 
The sawtooth lattice can be created by suitably using a superlattice potential on top of the Kagome lattice which has been 
created recently~\cite{Jo2012} (see Fig.~\ref{fig:saw}). 

In order to obtain the signature of the SS phase we compute the momentum distribution $N(k)$ using Eq.~\ref{eq:mom}. 
As expected, in SF$_0$~(SF$_{\pi/2}$) $N(k)$ has a peak at $k=0$~($k=\pi/2$). However, an intriguing feature appears in the SS phase which 
shows, interestingly, peaks at $k=0$ and $k=\pi/2$ as shown in Fig.~\ref{fig:md}. We also plot the peak strengths as 
a function of $|t'|/t$ at $k=0$ and $k=\pi/2$ in Fig.~\ref{fig:md2} which shows the transition from SF$_0$~-~MI~-~SS~-~SF$_{\pi/2}$ for $U=3$ and $\rho=1$. 
Hence, the appearance of the SS phase may be directly monitored in time-of-flight experiments from the multi-peaked momentum distribution.



\section{Conclusions} 
\label{sec:Conclusions}

We have discussed a novel mechanism for the formation of lattice supersolids for the particular case of a sawtooth lattice.  
The mechanism is based on the selective population of un-frustrated plaquettes in the presence of frustrated and unfrustrated plaquettes, and 
hence we expect supersolids in other lattices fulfilling that property.
We have shown that the supersolid exists for a broad range of lattice fillings, including 
various commensurate fillings, in particular unit filling. Interestingly the supersolid may be revealed not only by in-situ measurements, but 
by monitoring the momentum distribution in time-of-flight measurements. Since frustrated lattices, in particular sawtooth, may be realized using state of the art 
techniques, our results open hence a new feasible path for realizing supersolids in existing experiments with ultra cold atoms in optical lattices, 
without the need of long-range interactions.


\begin{acknowledgments}
We thank T. Vekua for enlightening discussions. 
We acknowledge support by the Center QUEST and the DFG Research Training Group 1729. 
Simulations were performed on the cluster system of the Leibniz Universit\"at Hannover.
\end{acknowledgments}


\end{document}